# Defect detection in III-V multijunction solar cells using reverse-bias stress tests

A. Cano[a,*], I. Rey-Stolle[a], P. Martín[a], V. Braza[b], D. Fernandez[b], I. García[a,*]

[a] Instituto de Energía Solar, Universidad Politécnica de Madrid, Avda. Complutense 30, 28040, Madrid, Spain

[b] Departamento de Ciencia de los Materiales e IM y QI, Universidad de Cádiz, Puerto Real, 11510, Cádiz, Spain

E-mail addresses: aitana.cano.perez@upm.es (A. Cano), ivan.garciav@upm.es (I. García)

**Abstract:**

Reverse biasing triple-junction GaInP/Ga(In)As/Ge solar cells may affect their performance by the formation of permanent shunts even if the reverse breakdown voltage is not reached. In previous works, it was observed that, amid the three components, GaInP subcells are more prone to degrade when reverse biased suffering permanent damage, although they present an initial good performance. The aim of this work is, firstly, to study the characteristics of the defects that cause the catastrophic failure of the devices. For this, GaInP isotype solar cells were analysed by visual inspection and electroluminescence maps and submitted to reverse bias stress test. We find that specific growth defects (i.e. hillocks), when covered with metal, cause the degradation in the cells. SEM cross-section imaging and EDX compositional analysis of these defects reveal their complex structures, which in essence consist of material abnormally grown on and around particles present on the wafer surface before growth. The reverse bias stress test is proposed as a screening method to spot defects hidden under the metal that may not be detected by conventional screening methods. By applying a quick reverse bias stress test, we can detect those defects that cause the degradation of devices at voltages below the breakdown voltage and that may also affect their long-term reliability.

1. ## INTRODUCTION

GaInP/Ga(In)As/Ge triple-junction solar cells are currently the most mature and widely used technology for concentration photovoltaic (CPV) applications and space power. These devices can degrade when operating under reverse bias, what could occur, for example, when a solar cell is totally or partially shaded. When they operate in reverse and close to the breakdown region, the generated current increases exponentially [1], which can lead to the damage of the device. In practice, with the use of bypass diodes, this risk becomes irrelevant for cell operation. However, previous works have shown that subjecting cells to a reverse voltage can give us important information not only about the behaviour of the devices when reverse biased [2] but also about the existence of hidden defects in the devices [3].

This previous work consisted of applying a reverse bias stress test to Ge, Ga(In)As and GaInP isotype cells [2], which are equivalent to the top, middle and bottom subcells in a GaInP/Ga(In)As/Ge triple-junction solar cell. The experiments start with an initial measurement of the forward dark I-V curve. Then, as a reverse bias stress test, the reverse dark I-V curve is measured up to a certain value of reverse voltage. After this test, the forward dark I-V curve is measured again to check whether the devices have undergone any change in their performance. This process is repeated several times but increasing in each repetition the final reverse voltage of the reverse dark I-V curve measurement.

By applying this stress test some recurring phenomena were observed, especially on isotype GaInP solar cells. For example, reverse dark I-V curves were found to be erratic, and the devices soon degraded even at low levels of reverse bias (around -4 V), notably below the expected breakdown voltage. Also, before the catastrophic failure of the devices, sudden abrupt increases in the leakage current, or decreases in the shunt resistance, were observed in their reverse dark I-V curves.

A hypothesis was developed trying to explain the degradation at reverse bias of the isotype GaInP solar cells. The origin of leakage current can be associated to conduction through defects in the crystal structure, such as dislocations [4], which are always expected at low densities although the GaInP subcells are grown lattice matched to the substrate with a high degree of crystalline perfection. These crystal defects can show a rapid increase in electrical conductivity, possibly due to microplasmas [5], leading to a decrease in shunt resistance and an increase in leakage current.

In a subsequent study [3], a slightly different reverse bias stress test was applied to GaInP isotype cells, which will be explained in detail in section 2.2. as the same one is applied in this work. In this case, the reverse voltage applied is maintained for a few seconds, and electroluminescence (EL) mapping and microscope analysis were done before and after the catastrophic failure occurred, in order to locate the physical origin of the failure. The results pointed to a particular type of defect that, when covered with metal (either a grid finger or a busbar), was the cause of the failure of the devices during the reverse bias stress tests. This again puts the GaInP sub-cell in the spotlight, as it is the one in contact with the front grid metallisation in a triple-junction solar cell.

Furthermore, it was observed that even some devices with a good initial performance -i.e. a clean forward dark I-V curve-, could also degrade when reverse biased. Thus, the detection of these defects responsible for the degradation requires more than doing a forward I-V curve measurement.

In this work we aim to identify and characterise which defects are responsible for the premature and catastrophic failure of GaInP devices and to understand how failure occurs. To have some statistical significance, we increase the number of devices studied and perform a thorough examination of the defects present during the processing of the devices. Furthermore, we notice that some otherwise healthy devices showing high quality forward characteristics fail prematurely under reverse bias, which indicates that reverse bias stress test can be used to reveal the presence of defects.

## 2. EXPERIMENTAL

### 2.1. Devices

The devices submitted to a reverse bias stress test have the semiconductor structure depicted in Fig. 1, which has been grown in a metalorganic vapour phase epitaxy (MOVPE) reactor [6]. All tested devices belong to the same wafer and were thus grown at the same time. Their structure corresponds to an isotype GaInP top cell, i.e. this cell operates in a similar way to the GaInP subcell on a triple-junction solar cell.

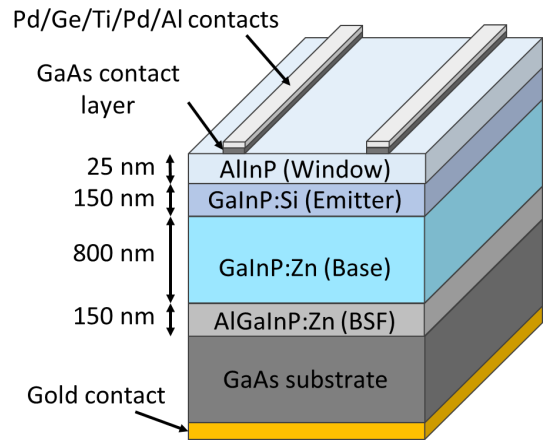

Fig. 1. Semiconductor structure of the devices submitted to the stress test.

Once the semiconductor structure is grown the devices are fabricated. For this purpose, in addition to mesa etching and defining devices with an area of 3x3 mm$^2$, the front and rear metal contacts are deposited. The front grid metallisation consists of a stack of Pd/Ge/Ti/Pd/Al deposited by electron-beam evaporation [7] and submitted to an annealing process. In addition, this front grid has an inverted-square shape [8], with a high finger density, which is suitable to minimise the series resistance in CPV applications. On the other hand, metal electroplated gold was deposited creating a blanket contact at the rear side.

### 2.2. Reverse Bias Stress Test

The reverse bias stress test applied in this work is schematically presented in Fig. 2. First, an initial characterization is done, including a forward dark I-V curve measurement, an EL intensity map and a visual analysis under the microscope. Moreover, before this characterisation and the processing of the devices took place, growth defects were located and mapped throughout the wafer, so those eventually covered with metal could be identified.

In order to perform this initial characterisation, a Leica DM1750 microscope was used for visual inspection, and a Keysight B2902A source meter unit was used for electrical measurements. These electrical measurements were performed ensuring a temperature of 25 °C in the device by using the 274 TECMount peltier connected to the 5305 TECSource temperature controller, both from Arroyo Instruments. Finally, the EL maps were taken using a Dino-Lite AM4013MLT camera.

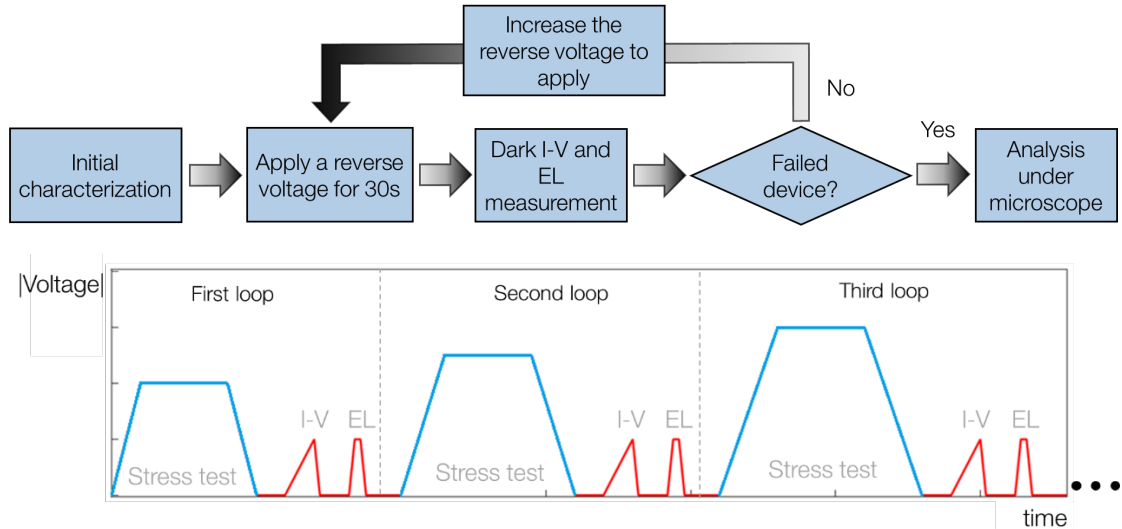

Fig. 2. Scheme of the stress test applied (top image). Representation of the voltage variation in absolute values during the test: in blue the negative voltages and in red the positive ones (bottom image).

This initial characterisation is followed by the stress test, in which a reverse voltage is applied for 30 seconds. In this way, we can study whether degradation occurs when a certain reverse voltage level is reached, or when this voltage is maintained for a certain time. Additionally, this voltage is not applied and released abruptly, but by decreasing and increasing the voltage progressively with a ramp (variation of -2 V/s) from 0 V to the target reverse voltage and vice versa. Thus, we can ensure that the degradation is caused by the level of reverse voltage applied and not by the sudden change in bias.

After the stress test the forward dark I-V curve and the EL are measured again. If the device has not failed, the stress test is repeated, but increasing the reverse voltage applied for 30 seconds (see the graph in Fig. 2). This process is repeated as many times as needed to cause the catastrophic failure of the device, or until a safety limit voltage is reached. This voltage limit was set at -20 V, a value that an undamaged device should withstand without degradation considering that the theoretical breakdown voltage of these devices should be around -(25-28) V. This value was calculated using eq. (1) -extracted from [9]- and considering a base layer doping level around $N=10^{17}$ cm$^{-3}$, a bandgap around 1.85 eV and their corresponding uncertainties due to the growth process.

$$V_{BD} \approx 60 \cdot \left(\frac{E_g}{1.1\ \text{eV}}\right)^{\frac{3}{2}} \left(\frac{N}{10^{16} \text{cm}^{-3}}\right)^{-\frac{3}{4}} V \qquad (1)$$

Finally, those devices that catastrophically fail (i.e., they become totally shunted) during the stress test before reaching the limit voltage, are analysed under the microscope trying to find out what has caused the failure. In addition, in order to understand the morphology of the defects and try to explain why they cause the degradation, Scanning Electron Microscopy (SEM) and Energy Dispersive X-ray (EDX) analyses were obtained in an analytical focused ion beam (FIB)-SEM system, FEI Scios™ 2 Dual Beam™ [10] [11], as will be further explained in section 2.3.2.

### 2.3. Results

#### 2.3.1. Phenomenology of the failure

By submitting the devices to the stress test described in the previous section, it is possible to understand better the failure mechanism of the devices when reverse biased. It was observed that our devices fail when a certain reverse voltage level, lower than the limit voltage, is reached during the stress tests, but only when there is a metal-covered defect in the device. In particular, a hillock growth defect which will be described in more detail in section 2.3.2.

Starting with the previous characterisation of the devices, in Fig. 3 the initial forward dark I-V curves of the devices submitted to the stress test is depicted. The blue curves correspond to the devices that withstand the stress test applied later on without degrading, and the orange curves correspond to the devices that do not pass the reverse bias stress test, and therefore, catastrophically fail during it. As expected, all devices that initially have a low shunt resistance, and therefore, a high current at low values of voltage, do not pass the stress test. However, focusing on the devices that have a good initial forward I-V curve, three of them fail at a reverse voltage below the breakdown voltage. This confirms the findings of previous work: a forward dark I-V curve measurement is not enough to predict the failure of the devices during the reverse bias stress test.

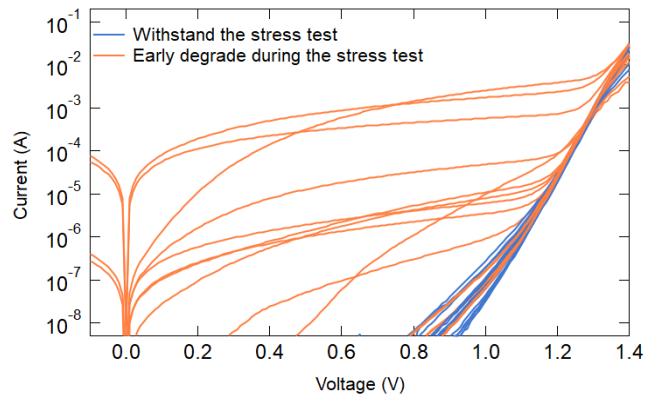

Fig. 3. Initial forward dark I-V curves of the GaInP devices submitted later on to the stress test.

The voltage level at which the failure occurs varies between -7 and -20 V. This value is not related to the initial state of the device, but there seems to be a tendency for devices with an initially low shunt resistance to fail at lower reverse biases. Moreover, although the reverse voltage is applied for 30 seconds, the failure occurs just as the voltage is applied, so it seems that it is the level of reverse voltage, and not the time it is applied, that causes the failure. However, we cannot discard the possibility that the failure appears also if a lower reverse voltage is applied for longer than 30 seconds [12].

While the devices are subjected to a reverse voltage for 30 seconds, current data is taken every 3 ms, so that it is possible to monitor whether the device has suffered any damage, even before measuring its forward dark I-V curve. As can be seen in Fig. 4, the current remains almost constant around an average value which increases with the applied voltage. This increase is more or less pronounced depending on the initial characteristics (shunting) of the device. For higher values of current, and as we approach the value at which the device fails, more abrupt variations in the current over time can be seen. These variations are present especially in devices with a poor initial forward dark I-V curve, and may indicate the existence of microplasmas [2]. Furthermore, at the voltage at which the failure occurs, the current increases abruptly to a value of 0.5 A (see graph on the right at Fig. 4), which is the value established as the protection or limit current, indicating the presence of a short circuit or a considerable decrease in the shunt resistance.

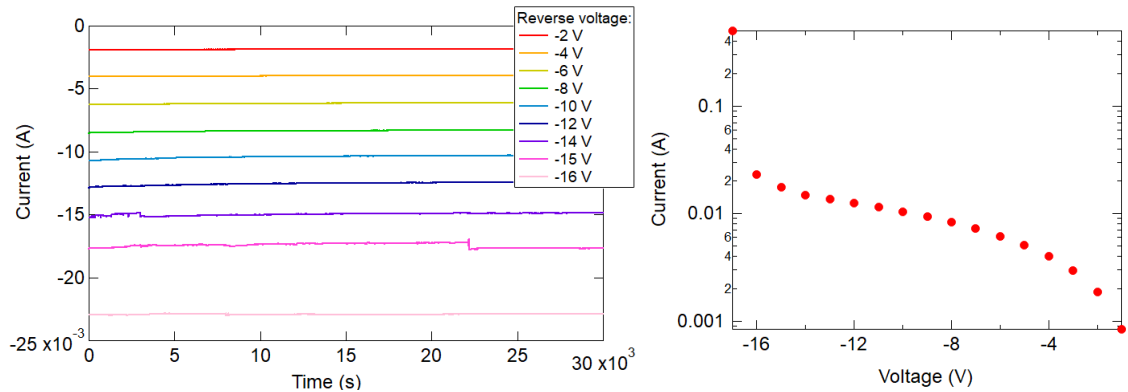

Fig. 4. Variation of the current with the time and the reverse voltage applied during the reverse bias stress test (graph on the left). Representation of the average value of the current measured for 30 seconds versus the reverse bias voltage applied (graph on the right).

Most devices with a leaky initial I-V curve, i.e. with a low shunt resistance, show changes in the I-V curve after applying reverse voltages lower than required for failure, making it predictable. However, for some devices, especially those with a good initial I-V curve, no change is detected in the I-V curve until it catastrophically fails, and when failure occurs, the current increases abruptly. In the forward dark I-V curve measured afterwards, it can be seen how the shunt resistance also decreases abruptly, so much so that it could be considered that the device is practically short-circuited (see Fig. 5).

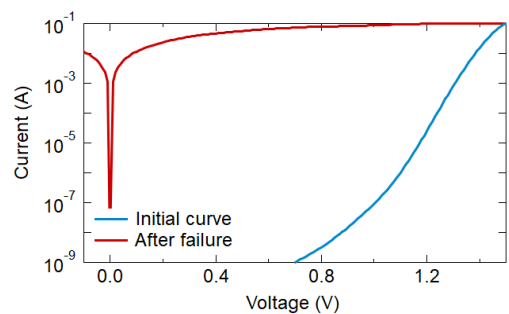

Fig. 5. Forward dark I-V curve measured before starting the reverse bias stress test and after the catastrophic failure of the device.

However, irrespective of the reverse voltage level at which the devices fail, after the failure, burnt and melted regions are observed right where a metal-covered defect is located. These burnt regions look different depending on whether the defect is located under a grid finger or a busbar. First, when the defect is under a finger, these regions appear along the entire finger (see Fig. 6), and not only on the defect. Thus, the failure of the device can be confirmed not only by measuring the forward I-V curve but also by doing EL mapping, in which the entire finger appears darkened (see Fig. 6), or by analysing it under the microscope.

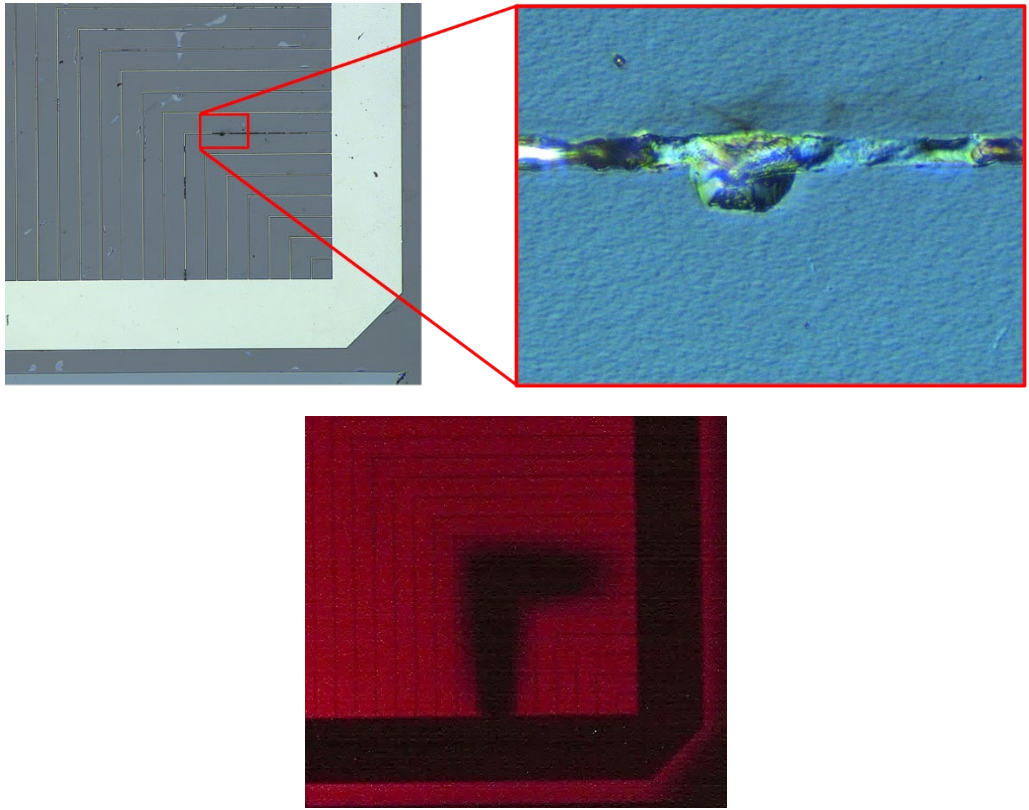

Fig. 6. Appearance of burnt and melted metal regions in a finger after the catastrophic failure where there is a defect covered with metal (top image). The same finger completely shadowed in EL mapping (bottom image).

As shown in Fig. 7, if the defect is under the bus the failure mechanism is basically the same: when examined under the microscope, burnt and melted metal regions appear where there was a metal-covered hillock and, again, a forward dark I-V curve reveals a device which is almost short-circuited. However, in this case, as the bus covers and masks the defect, the failure cannot be detected by EL mapping (see Fig. 7).

The devices that do not have growth defects covered with metal withstand reverse voltages up to the maximum -20 V tested, even those showing the same kind of hillock defects but located in the semiconductor area. These devices were subjected to higher reverse voltages up to -30 V. Eventually, these devices fail but the failure mechanism is not reverse breakdown, despite the reverse voltage used is close to or even higher than the theoretical breakdown voltage. Instead, dielectric breakdown is observed at some point of the edge of the device. An arc seems to appear between the bus and the edge of the devices causing a crack in the semiconductor and a burnt area in the bus (see Fig. 8). In this case, the failure can be located by looking at the EL measurement, where a black spot appears at the edge of the device in the vicinity of where the arc was formed (see Fig. 8). As in the previous cases, the electrical consequences of such failure are a sudden decrease of the shunt resistance, and therefore an almost short-circuited device.

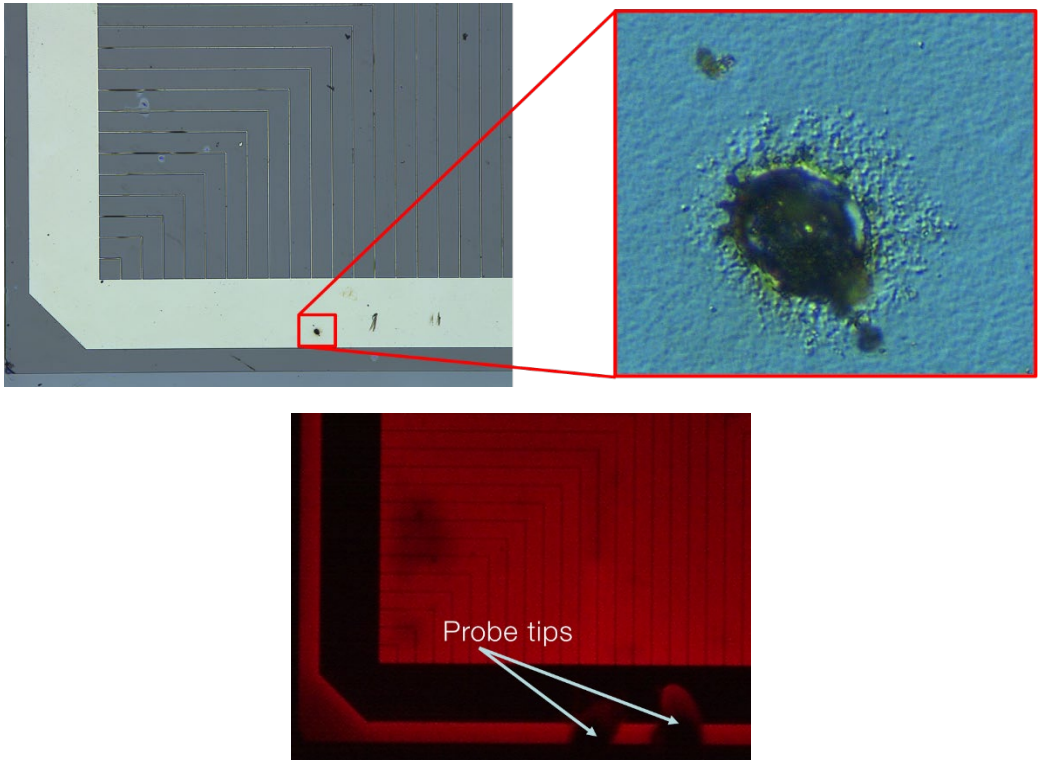

Fig. 7. Appearance of burnt and melted metal regions in the bus after the catastrophic failure where there is a defect covered with metal (top image). No change can be observed by doing EL mapping as the bus is completely shadowing the defect (bottom image).

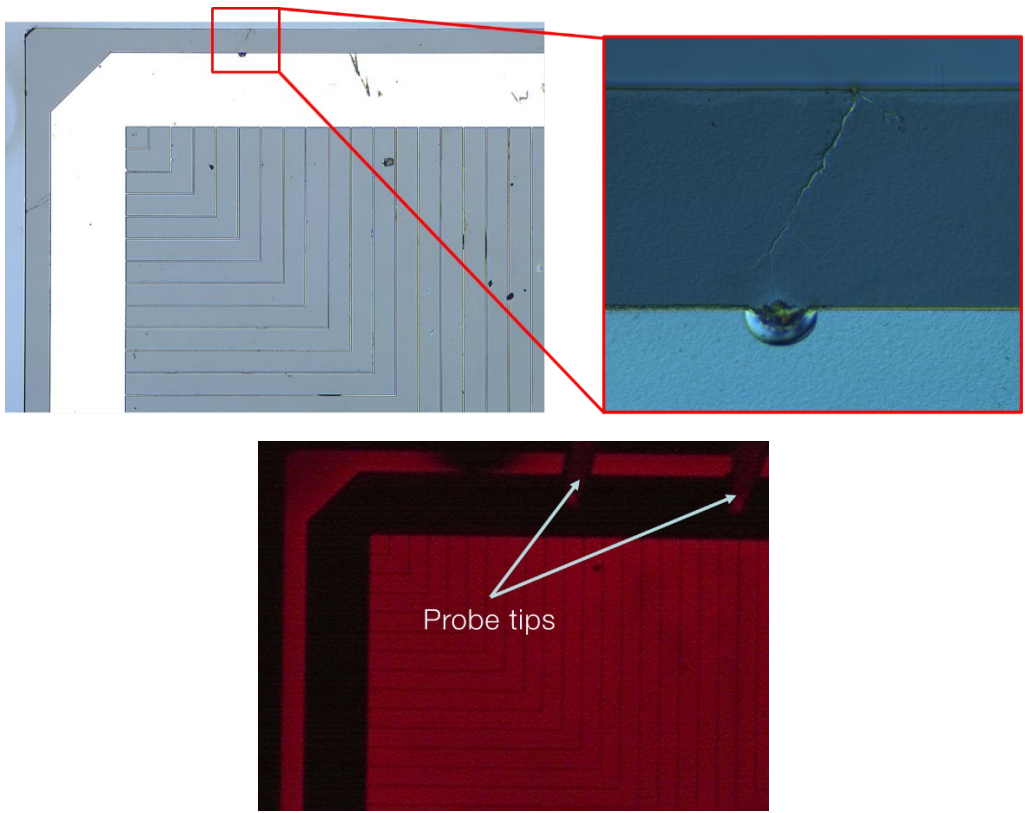

Fig. 8. Appearance of burnt and melted metal regions in the bus and cracks in the semiconductor due to a dielectric breakdown (top image). Black spot observed in the edge of the device by doing EL mapping (bottom image).

### 2.3.2. Physical mechanism of failure at the defects

Before processing the devices, the wafer was studied under the microscope, and a map of all defects with their precise locations on the wafer area was created. This made it possible to know which defects were under metal after the processing of the solar cells, and therefore which were most likely to fail during the reverse bias stress test. It was identified that in all our devices, the defects causing the degradation at reverse bias are a particular type of growth defects dubbed as *hillocks*, as the ones shown in Fig. 9, but only when they are covered with metal. Any other kind of defect observed on the wafers does not seem to affect the performance of the devices during the reverse bias stress tests performed.

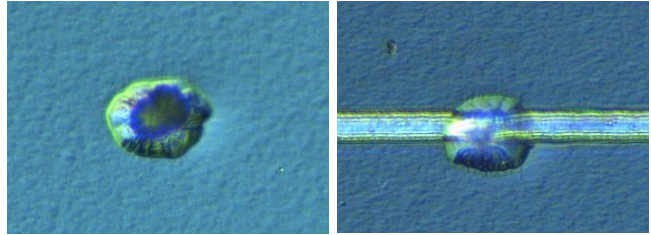

Fig. 9. Image under Nomarski microscope of a hillock without metal (left image) and partially covered with a metal grid finger (right image).

SEM and EDX analyses were done to a non-degraded defect partially covered with metal and some of the resulting images are shown in Fig. 10. First, with a top-image of the defect it was observed that the defect consists of a centrally located protrusion surrounded by a hillock. After covering the protrusion with platinum (in order to prevent it from detaching during the study), trenches were made by FIB etch processing that exposed the cross-section of the structure from the edge towards the centre of the defect. Thus, the morphology and composition of the defect could be studied by taking SEM images of different cross-sections of it.

The first image in Fig. 10 shows how the semiconductor structure is not affected in the vicinity of the defect. As we get closer to the centre of the defect, a bulge, mainly composed of GaInP, appears, as it is depicted in the second image. The existence of the protrusion also affects the semiconductor structure in its vicinity, where a hillock is formed. Finally, in third and fourth images, a particle appears in the core of the defect. It can be seen that the bulge is formed by growth of material on the surface of this particle, and under the particle there are voids as the deposited material does not reach these areas. Moreover, the colours of the particle clearly reveal that its composition is mainly GaAs with traces of other materials. The composition of the particle has been studied by EDX analysis and is available in the supplemental "SEM and EDX study performed". They show that the particle comprises the typical materials of a GaInP structure, indicating that it is originated by the formation of particles from the remaining of previous runs in the MOVPE reactor, as represented and detailed in Fig. 11. After detaching from somewhere in the reactor chamber and depositing in the GaAs wafer, the particle causes the formation of a cavities in the GaAs buffer layer and modifies the growth direction of the upper layers. This reveals higher index planes producing higher growth rates, which create a hillock in the surroundings of the particle. Moreover, the material being deposited also grows at the surface of the particle, with high growth rates, creating a bulge on it with a large thickness, compared to the thickness of the layers deposited on the wafer surface.

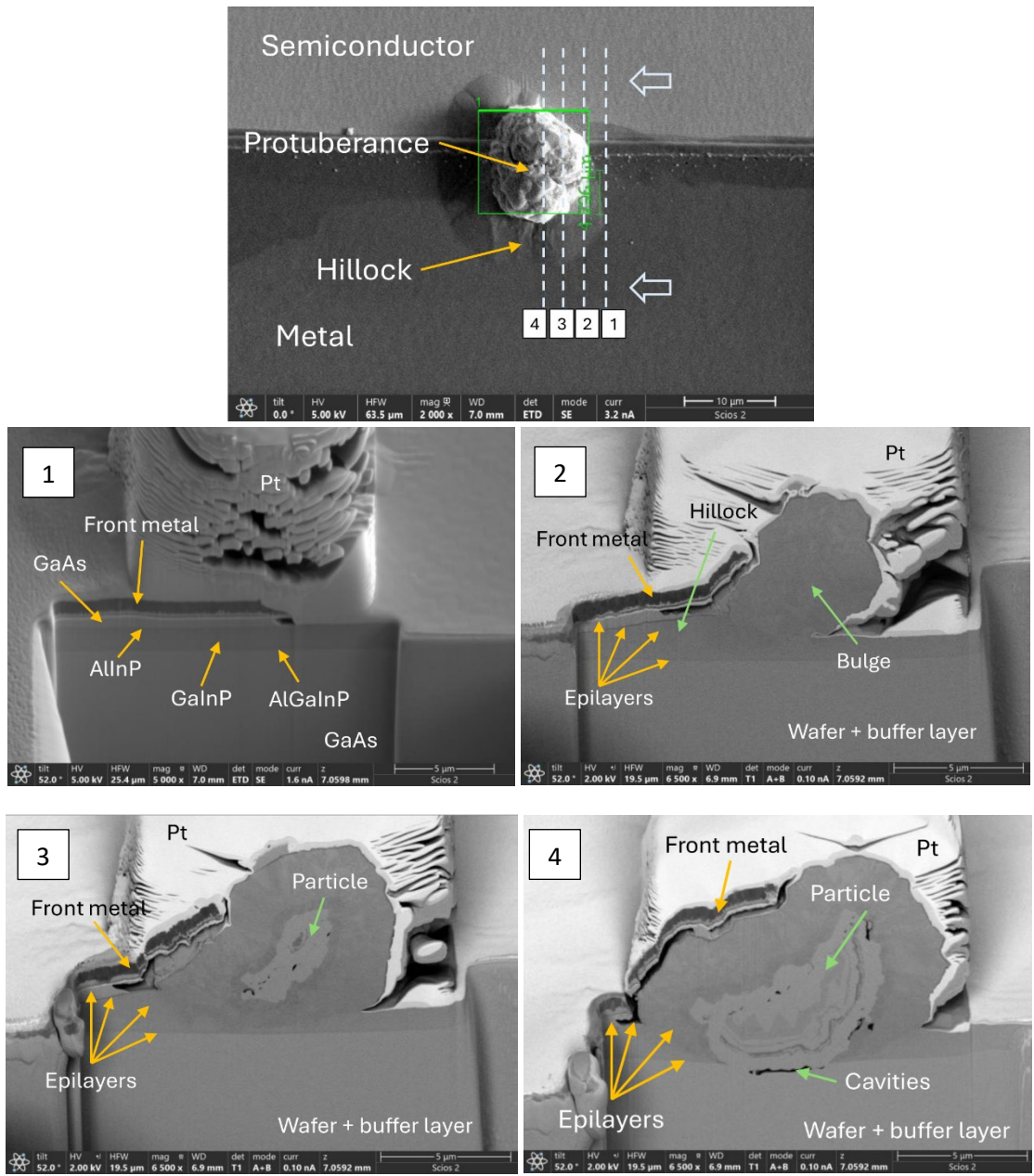

Fig. 10. SEM images of a defect partially covered with metal. Top and general view of the defect without platinum indicating the cutting planes of the cross-section images. (top image). Cross-section images from the outside to the inside of the defect (images numbered 1 to 4).

The morphology of the defects shown in Fig. 10 suggest the presence of leakage paths. The particle itself has p- and n-type layers, since it comes from previous depositions of materials with some doping. Also, during the growth of the semiconductor structure, as the upper layers adapt to the particle shape, areas with high mechanical stress and dislocations are formed. In addition, due to the non-uniformity of the GaInP layer thickness, there may be areas where the n-p junction is closer to the metal, and therefore more likely to cause a shunt as the applied reverse voltage increases.

Under reverse bias, most of the current will flow through the leakage paths created at the hillock. Note that the total resistance of these leakage paths is lower if there exist a metal region nearby or over the defect, facilitating the achievement of high leakage currents. High current densities lead to high power dissipation, and therefore increase of the temperature in small areas, as studied in silicon solar cells [13]. The higher the reverse voltage applied, the higher the current densities, power dissipated and temperature at the defect. Moreover, this thermal runaway process can lead to electromigration of the metal [14], forming a shunt large enough to cause the short-circuit of the device and the appearance of burnt and melted regions in the metal deposited on the defect responsible for the failure.

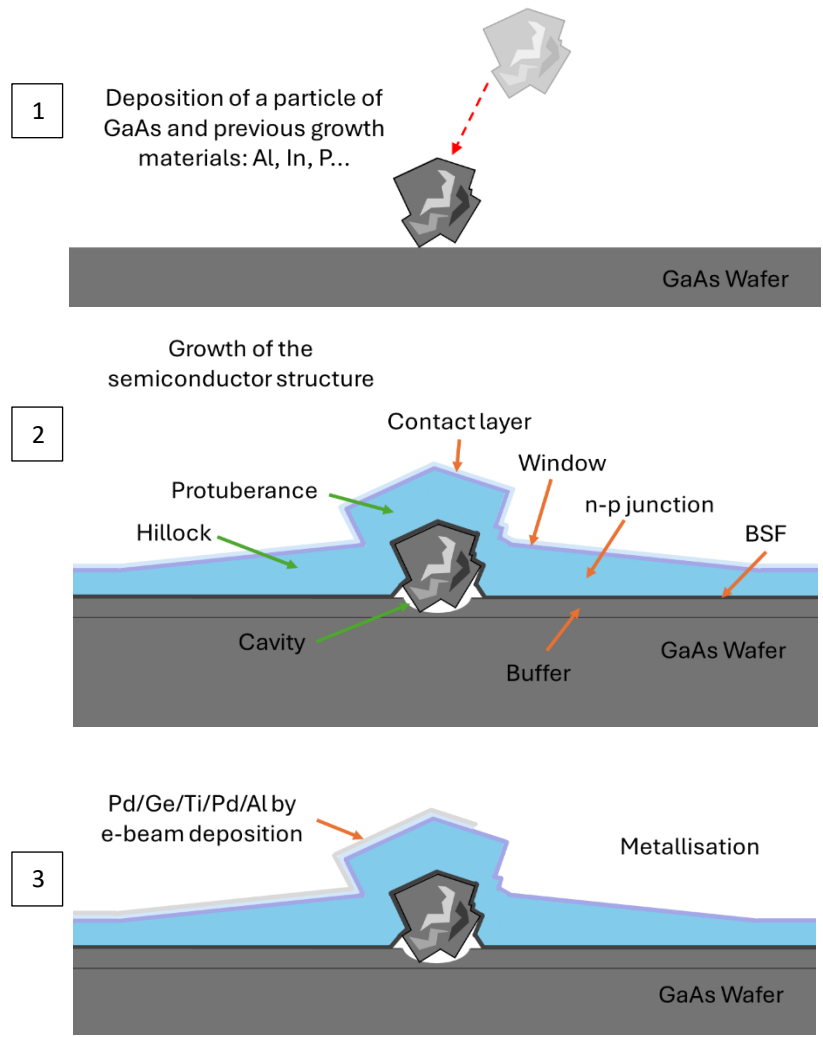

Fig. 11. Diagram explaining the process of defect formation. (1) Particle detached from the reactor chamber and deposited in the wafer. (2) Growth of the semiconductor structure and formation of the hillock, the bulge and cavities under the particle. (3) Metallisation.

## 3. REVERSE BIAS AS SCREENING METHOD

It has been observed that a forward dark I-V curve is not always enough to detect this kind of defect, as they may be only activated when the devices are reverse biased at a certain voltage level (as seen in Fig. 3). In addition, the defects may be completely hidden under the metal, thus visual methods such as microscope analysis or EL mapping may not be enough to detect them. It could be argued that the cells will never be exposed to such high reverse bias during operation and, hence, these hidden defects will never be activated. However, it is expectable that the defects could also affect the reliability of the devices under forward bias at a lower but steady pace. That is why it is important to be able to detect them in a screening process, which may not be always possible by using conventional methods.

Thus, we propose to use a reverse bias stress test to detect these defects, with the idea being to activate them and reveal their presence. This is why such reverse voltage level has to be sufficient to ensure that the defects are activated, but without reaching the breakdown voltage, e.g. 75-80% of the breakdown voltage, which in our case corresponds to a voltage around -18 to -22 V. Once this value is determined, it is sufficient to ramp up to that certain voltage level, maintain that voltage for a few seconds, and ramp down to 0 V (see Fig. 12), without the need to progressively increase the reverse voltage value and to analyse the device after each step.

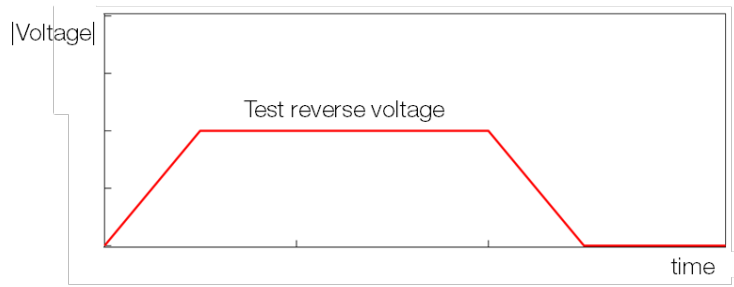

Fig. 12. Scheme of the voltage applied in absolute values in the reverse bias stress test as a screening method.

Adding this test to a screening process such as those included in CPV solar cell qualification standards, like IEC FDIS 62787, would not be a major effort, as it would require the same equipment as that used for an I-V curve measurement, so both measurements could be done at the same time. Compared to visual screening methods that are time consuming if not automated, this test can take only a few seconds. Furthermore, it does not pose a risk, since, as we have shown, healthy devices, i.e. those without defects, are not affected by high levels of reverse voltage provided they are kept below dielectric breakdown.

## 4. CONCLUSIONS

GaInP subcells submitted to a moderate reverse bias can catastrophically fail, way before reaching their breakdown voltage, regardless of presenting a *good* initial forward dark I-V curve.

In this work it was found out that hillock growth defects that end up covered, at least partially, with the front grid metallisation are the cause of failure at reverse bias, due to a process of thermal runaway and electromigration. A SEM and EDX study has shown that the origin of these hillocks is the presence of a particle on the wafer surface before the epitaxial growth starts, which creates an intricate structure that facilitates the appearance of leakage paths. The conduction of current through these leakage paths is enhanced when the front grid metal is in the proximity of the defect.

These defects may be completely hidden under the metal, so they may not be detected by visual inspection or EL mapping. Even though they seem to be only activated at reverse bias, which is unlikely to occur at significant levels in standard set-ups with bypass diodes, they can pose a significant reliability risk in the long term. That is why a new reverse bias stress test is proposed, complementary to the visual inspection and other tests included in qualifying standards for CPV solar cells like IEC FDIS 62787, to be applied during the screening process of solar cells.

**ACKNOWLEDGMENT**

The authors thank Luis Cifuentes and Jesús Bautista for technical assistance during device processing. This work has been partially supported by the Grant PID2021-123530OB-I00 funded by Ministerio de Ciencia e Innovación (MCIN/AEI/10.13039/501100011033). Part of the equipment used in this research for solar cell manufacturing was acquired through project LABCELL30 [grant number EQC2021-006851-P] with funding from the Spanish Ministerio de Ciencia e Innovación /Agencia Estatal de Investigación [MCIN/AEI 10.13039/501100011033] and the European Union "Next Generation EU"/PRTR", European Regional Development Fund (ERDF) "A way to make Europe" and by the Universidad Politécnica de Madrid through "Ayudas para la cofinanciación de infraestructuras de I+D+I (Programa Propio)".

**REFERENCES**

[1] P. A. Iles, H. I. Yoo, C. Chu, J. Krogen, and K.-I. Chang, "Reverse I-V characteristics of GaAs cells," in *IEEE Conference on Photovoltaic Specialists*, May 1990, pp. 448–454 vol.1. doi: 10.1109/PVSC.1990.111664.

[2] P. Martín, J. R. González, I. García, C. Algora, and I. Rey-Stolle, "Study of the reverse I–V in component subcells of III–V multijunction space solar cells," *Prog. Photovolt. Res. Appl.*, vol. 30, no. 5, pp. 481–489, 2022, doi: 10.1002/pip.3513.

[3] A. Cano, I. García, P. Martín, and I. Rey-Stolle, "Study of the Causes of Degradation of Space III-V Multijunction Solar Cells at Reverse Bias Operation," in *2023 13th European Space Power Conference (ESPC)*, Oct. 2023, pp. 1–5. doi: 10.1109/ESPC59009.2023.10298165.

[4] O. Breitenstein, J. P. Rakotoniaina, M. H. Al Rifai, and M. Werner, "Shunt types in crystalline silicon solar cells," *Prog. Photovolt. Res. Appl.*, vol. 12, no. 7, pp. 529–538, 2004, doi: 10.1002/pip.544.

[5] J. R. Gonzalez, I. Rey-Stolle, C. Algora, and B. Galiana, "Microplasma breakdown in high-concentration III-V solar cells," *IEEE Electron Device Lett.*, vol. 26, no. 12, pp. 867–869, Dec. 2005, doi: 10.1109/LED.2005.859626.

[6] G. B. Stringfellow, *Organometallic Vapor-Phase Epitaxy: Theory and Practice*. Elsevier, 1999.

[7] P. Huo and I. Rey-Stolle, "Al-based front contacts for HCPV solar cell," *AIP Conf. Proc.*, vol. 1881, no. 1, p. 040004, Sep. 2017, doi: 10.1063/1.5001426.

[8] A. R. Moore, "An optimized grid design for a sun-concentrator solar cell.," *RCA Rev.*, vol. 40, pp. 140–152, 1979.

[9] S. M. Sze, Y. Li, and K. K. Ng, *Physics of Semiconductor Devices*. John Wiley & Sons, 2021.

[10] P. E. Russell, "SEM Based Characterization Techniques For Semiconductor Technology," in *Spectroscopic Characterization Techniques for Semiconductor Technology I*, SPIE, May 1984, pp. 183–191. doi: 10.1117/12.939304.

[11] J. Melngailis, "Focused ion beam technology and applications," *J. Vac. Sci. Technol. B Microelectron. Process. Phenom.*, vol. 5, no. 2, pp. 469–495, Mar. 1987, doi: 10.1116/1.583937.

[12] N. C. Chen, Y. N. Wang, Y. S. Wang, W. C. Lien, and Y. C. Chen, "Damage of light-emitting diodes induced by high reverse-bias stress," *J. Cryst. Growth*, vol. 311, no. 3, pp. 994–997, Jan. 2009, doi: 10.1016/j.jcrysgro.2008.09.123.

[13] J. W. Bishop, "Microplasma breakdown and hot-spots in silicon solar cells," *Sol. Cells*, vol. 26, no. 4, pp. 335–349, Sep. 1989, doi: 10.1016/0379-6787(89)90093-8.

[14] P. S. Ho and T. Kwok, "Electromigration in metals," *Rep. Prog. Phys.*, vol. 52, no. 3, p. 301, Mar. 1989, doi: 10.1088/0034-4885/52/3/002.

# SEM and EDX study performed

This supplemental includes the images resulting from the Scanning Electron Microscopy (SEM) and Energy Dispersive X-ray (EDX) analyses done by FEI Scios™ 2 Dual Beam™ at the Universidad de Cádiz, as well as a brief description of the process followed.

First, in Fig. 1 you can see a general top-view image of the analysed hillock growth defect. The defect is partially covered with metal (Pd/Ge/Ti/Pd/Al front metal grid). Then, in Fig. 2, the same defect is shown, but with platinum deposited to prevent the particle from detaching during the study.

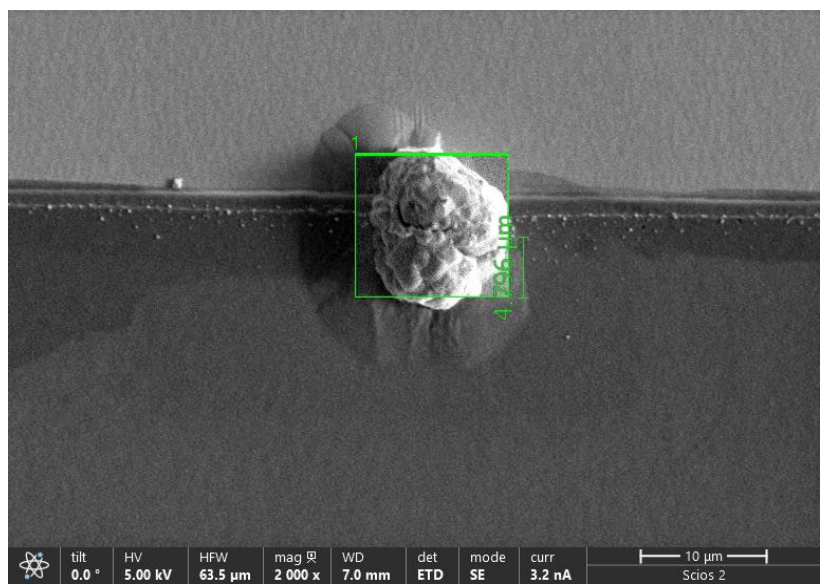

Fig. 1. Top-view of the defect under study.

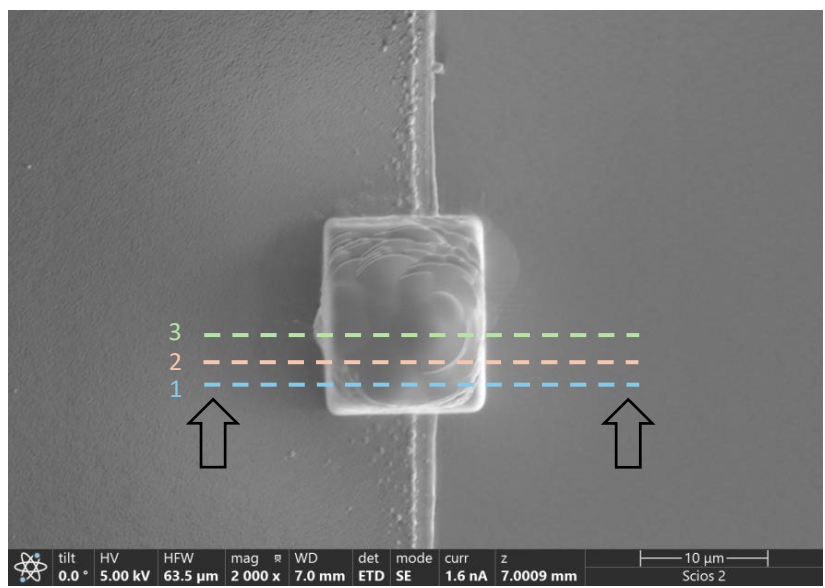

Fig. 2. Same defect than in Fig. 1 but with platinum deposited.

Focused ion beam (FIB) etch processing is then used to create trenches (as the one shown in Fig. 3) from the edge towards the centre of the defect, exposing the cross section of the structure. In Fig. 4, some resulting SEM cross-section images are included. To understand which part of the defect each image corresponds to, in Fig. 2, dashed lines corresponding to the approximate position of each cross-section image have been drawn. Arrows have also been included to understand from which side the defect is being viewed.

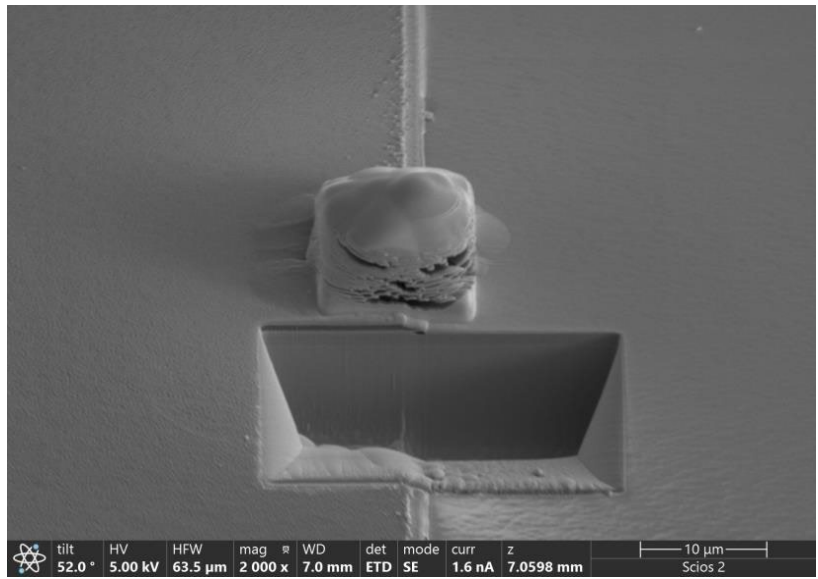

Fig. 3. Trench done by FIB.

For each of the cross-section images included in Fig. 4, the composition of the structure has been analysed by EDX. The maps extracted for each of the materials analysed by EDX are included in Fig. 5, Fig. 6 and Fig. 7, as well as a clarification of the area being analysed in each of the cross-section images of Fig. 4.

Fig. 5 shows how each layer is dominated by the corresponding materials of a GaInP mono-union deposited on a GaAs substrate, with GaAs contact layer, AlGaInP BSF layer, AlInP window layer and Pd/Ge/Ti/Pd/Al front metal contact. The same materials can be detected in Fig. 6, as we have the same structure but following the shape of the defect.

Finally, in Fig. 7, we can see that the particle causing the defect formation is composed mainly of gallium and arsenic. However, as can be seen in the SEM image, there are layers inside the particle of other materials, whose composition cannot be detected with EDX. That is why, new images are taken by zooming in on the particle, which are included in Fig. 8. As can be seen, the layers inside the particle present gallium, aluminium, phosphorus and indium signals. These materials correspond to the ones used in the GaInP, GaAs, AlGaInP and AlInP typical layers deposited during the growth of triple-junction solar cell structures in the reactor used.

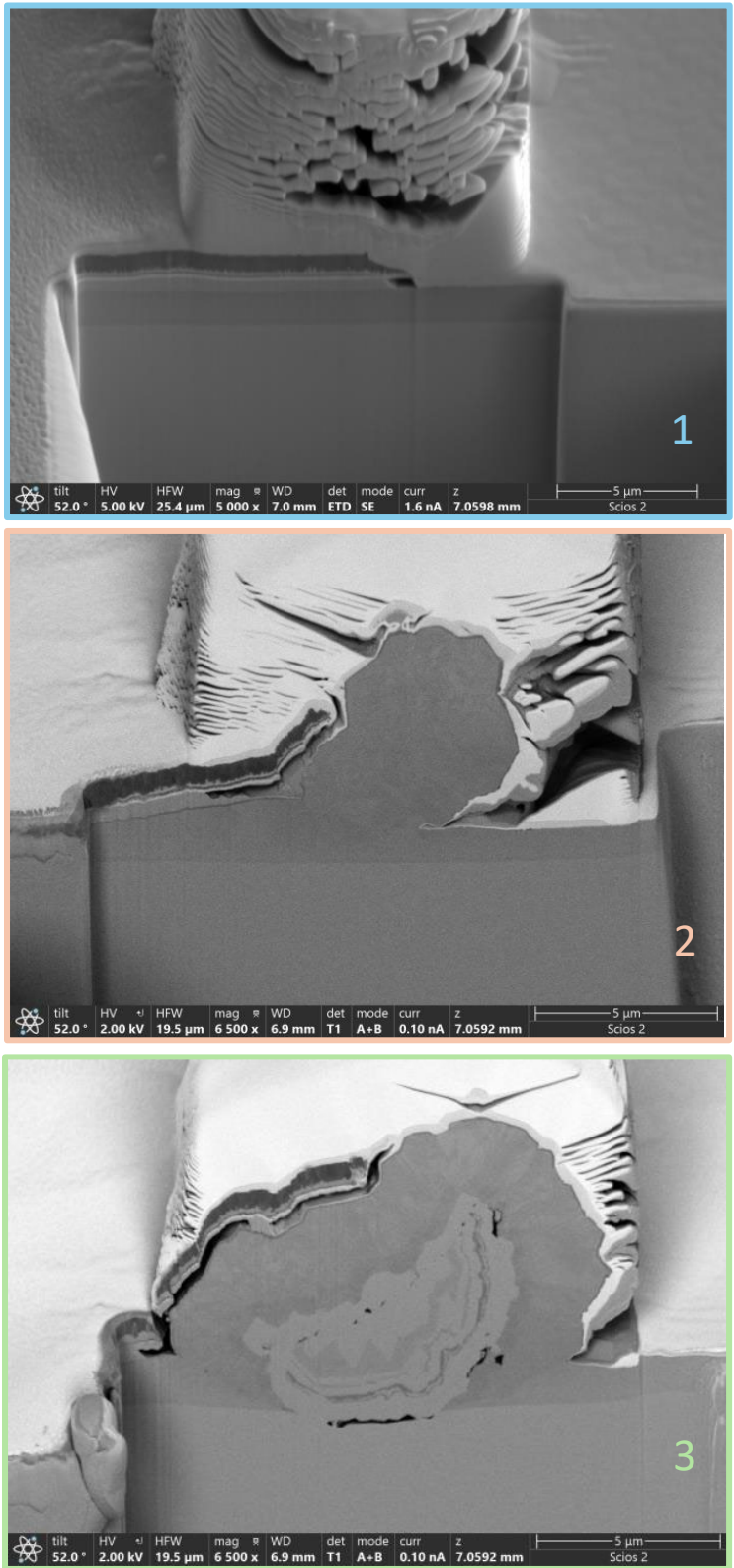

Fig. 4. SEM cross-section images from the edge towards the centre of the defect.

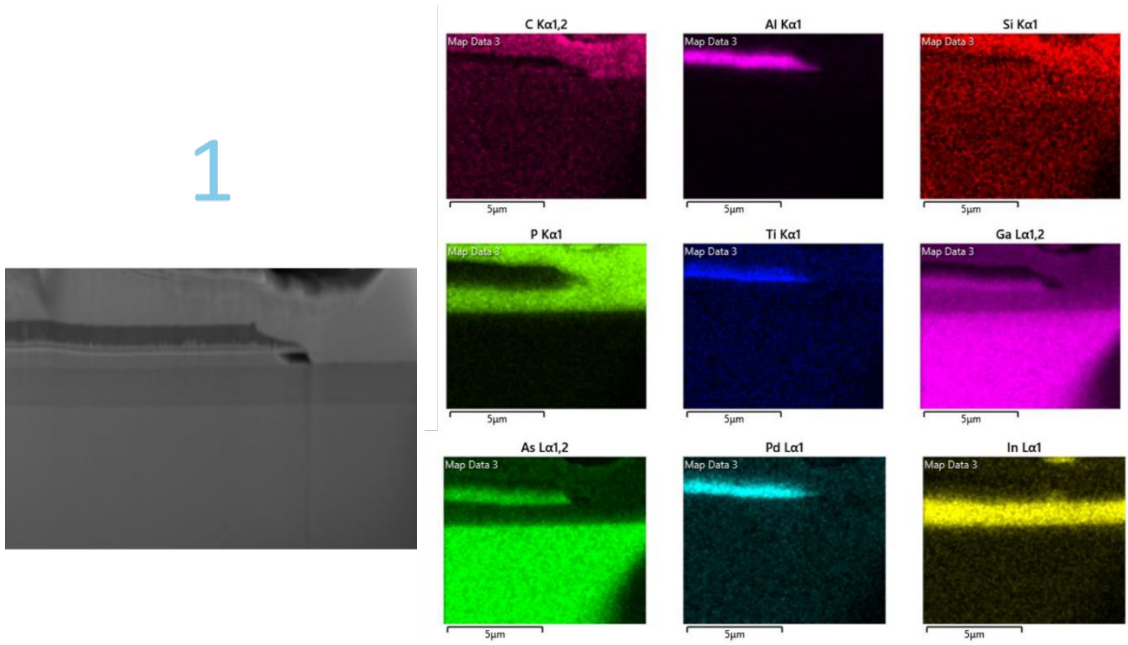

Fig. 5. EDX maps with the composition of part of the structure of image 1 in Fig. 4.

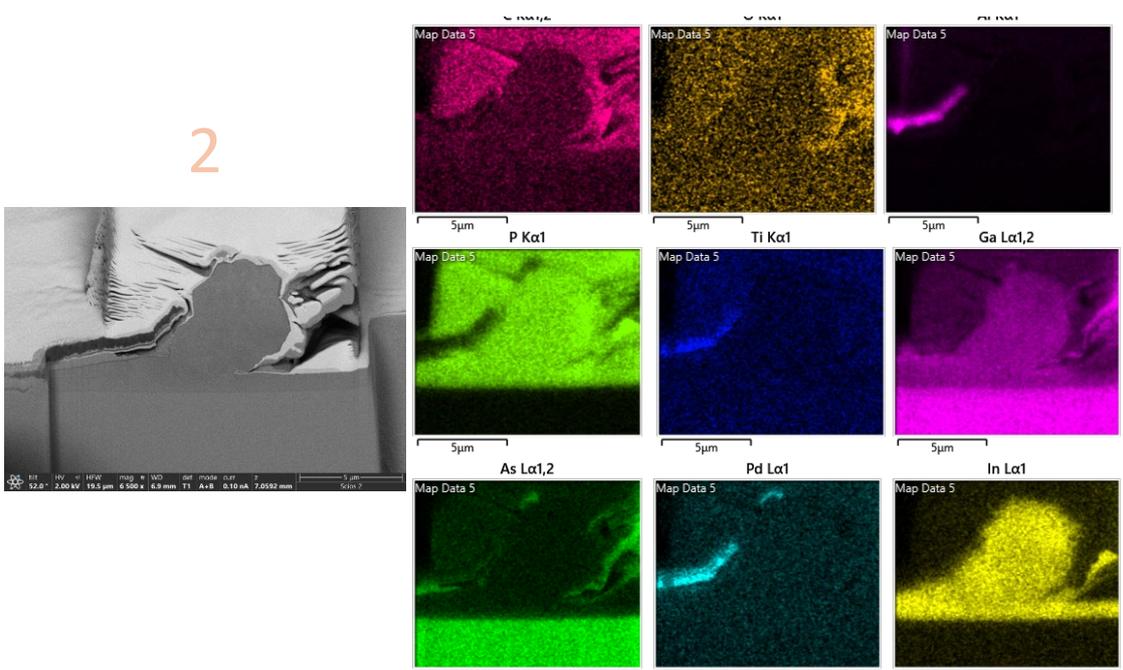

Fig. 6. EDX maps with the composition of part of the structure of image 2 in Fig. 4.

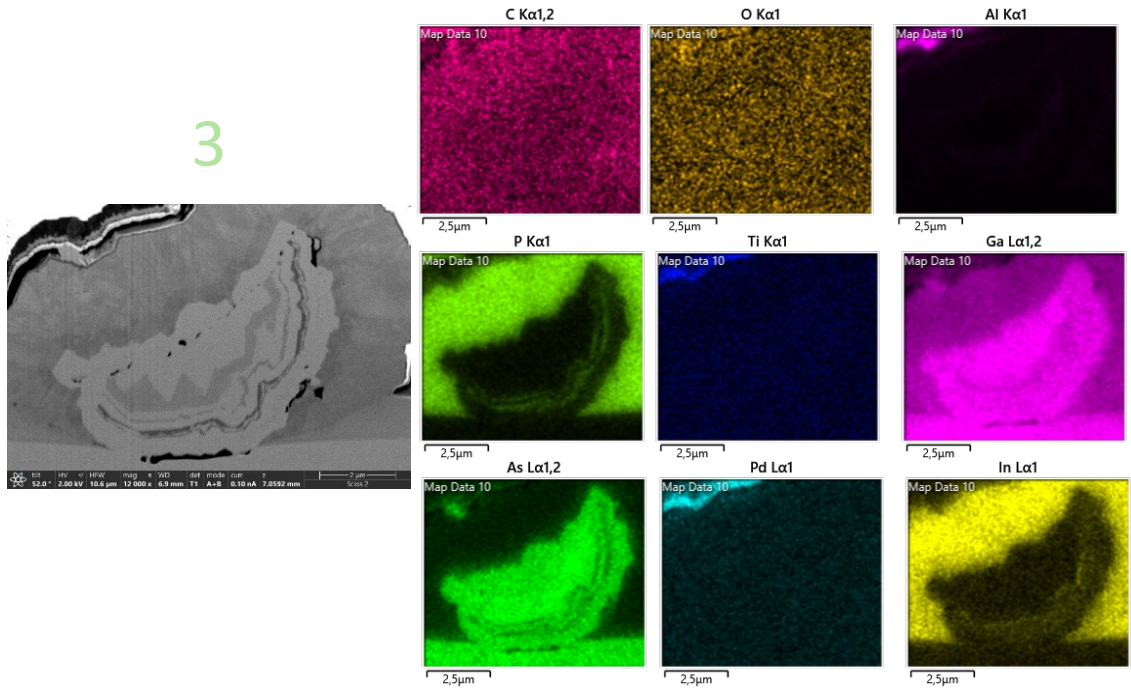

Fig. 7. EDX maps with the composition of part of the structure of image 3 in Fig. 4.

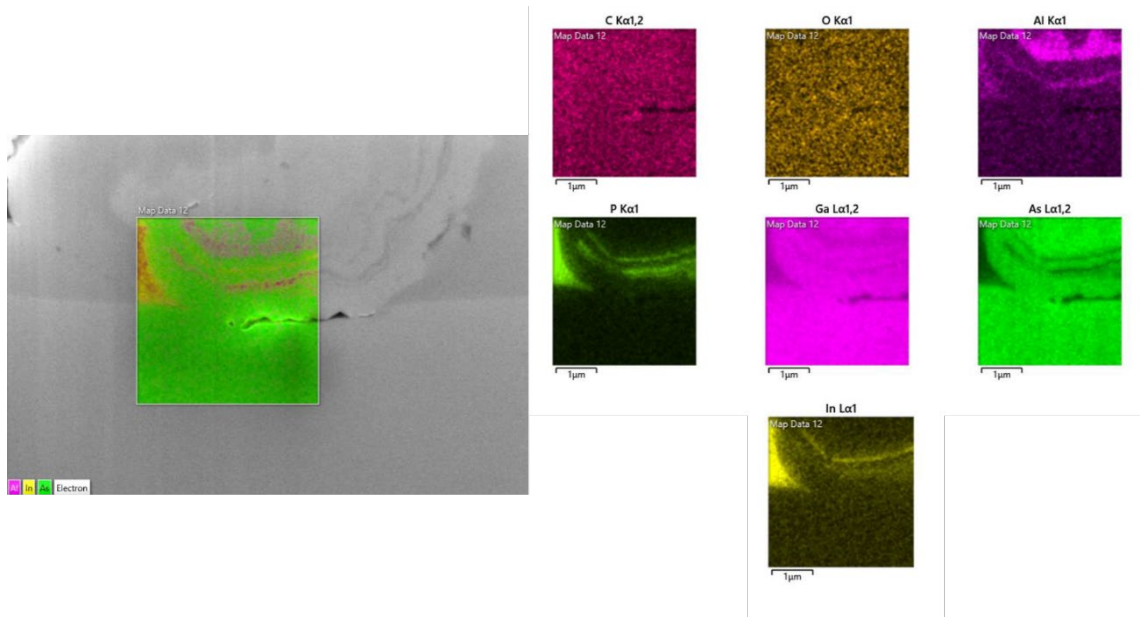

Fig. 8. EDX maps taken to see the composition of the particle.